\documentclass[journal]{IEEEtran}
\usepackage{balance}
\ifCLASSINFOpdf
\usepackage[pdftex]{graphicx}
\else
\fi
\usepackage{graphicx}

\begin{document}
\title{Wide Band Time-Correlated Model for Wireless Communications under Impulsive Noise within Power Substation}

\author{Fabien~Sacuto,~\IEEEmembership{Student member,~IEEE}
        Fabrice~Labeau,~\IEEEmembership{Senior member,~IEEE}
        and~Basile~L~Agba,~\IEEEmembership{Member,~IEEE}
\thanks{F. Sacuto is with the Department
of Electrical and Computer Engineering, McGill University, Montreal,
QC, H3A 0E9 Canada (email: fabien.sacuto@mail.mcgill.ca).}
\thanks{F. Labeau is with the Department
of Electrical and Computer Engineering, McGill University, Montreal,
QC, H3A 0E9 Canada (email: fabrice.labeau@mcgill.ca).}
\thanks{B.L. Agba, is with ASM, IREQ, Varennes,
QC, J3X 1S1 Canada (email: agba.basile@ireq.ca).}}

\markboth{IEEE TRANSACTIONS ON WIRELESS COMMUNICATIONS}%
{Shell \MakeLowercase{\textit{et al.}}: Bare Demo of IEEEtran.cls for Journals}

\maketitle

\begin{abstract}


The installation of wireless technologies in power substations requires characterizing the impulsive noise produced by the high-voltage equipment. Substation impulsive noise might interfere with classic wireless communications and none of the existing models can reliably represent this noise in wide band. Previous studies have shown that impulsive noise is characterized by series of damped oscillations with the amplitude, the duration and the occurrence times of the impulses that are random. All these characteristics make this noise time-correlated and the partitioned Markov chain remains an efficient model that can ensure the correlation between the samples. In this study, we propose to design a partitioned Markov chain to generate an impulsive noise that is similar to the noise measured in existing substations, in time and frequency domains. We configure our Markov chain to produce the impulses with the damped oscillation effect, then, we determine the probability transition matrix and the distribution of each state of the Markov chain. Finally, we generate noise samples and we study the distribution of the impulsive noise characteristics. Our Markov chain model can replicate the correlation between the measured noise samples; also the distributions of the noise characteristics are similar in the simulations and the measurements.
\end{abstract}


\IEEEpeerreviewmaketitle

\section{INTRODUCTION}

\IEEEPARstart{W}{ireless} technologies used in sensor networks must ensure reliable transmission of data, which requires some knowledge about the characteristics of the environment where the communication is performed. In a smart grid context, energy providers plan the installation of wireless sensor networks within power substations, where the channel is different from what the classic wireless communications are expecting. The electromagnetic environment of power substations has already been studied in several research areas, such as electromagnetic compatibility (EMC)~\cite{pakala1,pakala2}, power equipment maintenance~\cite{bart} or communications~\cite{moore1,moore2}, and the main conclusions coming from these studies are that impulsive noise is mainly created by partial discharges, corona noise and electrical arcs, that are all hosted by high-voltage equipment such as transformers, bushings, power lines, circuit-breakers and switch-gear.\

The partial discharges are electrical phenomenons located within different high-voltage pieces of equipment, with occurrences and lifetimes that are unknown. More precisely, a partial discharge is an electric current occurring in a microscopic area of an insulator between two conductors under different electrical potentials. An insulator usually becomes less efficient in some microscopic areas due to aging, then, when an electrode drives an electrical field above a critical value of the insulator degradation, a partial current occurs~\cite{bart}. It can be located in the oil of a transformer~\cite{app1}, on a bushing surface, on power lines and it is strongly related with the feeding voltage of the equipment~\cite{taiwan}. The electromagnetic noise coming from partial discharges evolves with the weather~\cite{hikita}, the voltage and the nature of the material of the equipment~\cite{taiwan}; moreover in a substation containing a lot of equipment, the location of impulsive noise sources is hard to predict and hardly spottable, which makes the study of this noise difficult.\

Our previous works have studied the noise in power substation for the 800 MHz - 2.5 GHz band that contains many of the wireless frequencies carriers and the power of the impulses was greater than the sensitivity range of classic wireless receivers~\cite{taiwan,cigre}. If the impulses are powerful enough to be detected by any commercial wireless devices, then, an increase of the impulse duration, amplitude and occurrence will increase the impulsive noise power, which is more likely to interfere with the wireless communications and to disturb the electrical network in smart grid applications. Other studies of wireless communications in power substations have confirmed the threat of impulsive noise on the reliability of the transmission~\cite{WC1, WC2, WC3, WC4}. So far the main propositions to reduce the interferences are either to increase the transmission power or to choose a safe path for the sensors disposition (safety distance from the equipment or antenna orientation).\

In order to adapt digital wireless communications to the interferences within substations, the receivers must be implemented with a reliable sampled model of impulsive noise; moreover some classic wireless technologies, such as Wi-Fi, ZigBee, Bluetooth, Wimax, LTE, must be considered, which requires a wide band model. From our previous studies~\cite{cigre,taiwan}, we have observed that impulsive noise in power substations is composed of series of damped oscillations, which implies that the samples of the impulses are correlated, so an appropriate noise model should replicate such a correlation between the samples. The existing measurement campaigns in substations~\cite{cigre} have also revealed that impulsive noise has characteristics, i.e the duration, the amplitude and the occurrence times of the impulses, that have particular distributions; any model should thus be able to replicate similar distributions.\

This paper introduces, in Section II, general concepts about impulsive noise and the different methods to represent it. In Section III, we present the partitioned Markov chain model and the configurations it can have to represent impulsive noise, including the implementation of damped oscillations. Thereafter, in Section IV, we describe the method to estimate the parameters of our model, more specifically, the probability transition matrix and the distributions of the samples value of the Markov chain states. Finally, we will present the results and discuss the performances of our model in the Section V.


\section{IMPULSIVE NOISE}

\subsection{Theory of impulsive noise}
A general approach to understand impulsive noise is to consider it being ruled by a process that switches from a background noise to another noise for a short duration~\cite{vaseghi}. Choosing a model of impulsive noise that represents an environment requires identifying the noises that are switched, the sample distributions and the switching rule.\

Depending on the sampling frequency used, the impulses can be considered as one sample or a burst of samples. Most models consider the samples of impulsive noise being i.i.d., which simplifies the generation of the noise and the parameter estimation. The Bernoulli-Gaussian~\cite{vaseghi}, other Gaussian mixtures and Middleton class A, B and C models~\cite{midd1} consider the noise samples to be i.i.d.\

Different approaches attempt to explain the behavior of impulsive noise in order to find the best way to represent it. So far, Middelton has proposed different models (Class A, B and C)~\cite{midd1} to characterize impulsive noise at the receiver. Each Middleton model attempts to represent simultaneously several kinds of impulsive environments, which results in approximating the noise characteristics, such as the shape and duration of the impulses. Although the Middleton models are often used in communication studies, they might be inappropriate for representing the substation environment. The Class-A model might be able to generate samples of impulsive noise, but so far, there is no reliable way to generate samples that produce a noise spectrum similar to the measurements performed in~\cite{taiwan,cigre} and that emulate the time-correlated nature of the interferences in power substation~\cite{cigre}.\

Another example of impulsive noise model is the output of an impulse-shaping filter excited by a random binary sequence modulated in amplitude~\cite{vaseghi}; the sequence of impulsive noise samples is composed of short impulses with random occurrence times and amplitudes, but the shape of the impulses remains unchanged, which is unlikely to be observed.\

The pattern of the impulses is time-varying, so the impulsive noise contains impulses with different durations, different rise times and fall times, and different oscillations. When representing impulsive noise in wide band, the difficulty is to replicate the waveforms of the impulses observed in measurements in order to have a similar spectral content during simulations.
\subsection{Noise characteristics observed in existing measurement campaigns}
The existing measurement campaigns in air-insulated substations were first performed in the spectral domain~\cite{pakala1,pakala2}, but with the improvement of measurement equipment such as antennas and digital oscilloscopes, today the measurements of impulsive noise can be performed in time-domain for a large bandwidth and with an accurate resolution. The time-domain measurements in power substations show a background noise with short oscillations occurring randomly~\cite{moore2}, which confirms that impulsive noise is time-correlated and that it mainly characterizes the RF environment in high-voltage substations. Also, from measurement campaigns that we have led in substations owned by a Quebec utility in Canada, Hydro-Quebec~\cite{taiwan,cigre}, we have noticed that a correlation might exist between the duration and the amplitude of the impulses. The observation of these two variables have provided a Pearson correlation coefficient of 0.85, which confirmed our assumption: the larger the impulses are, the longer they last.
\begin{figure}[h]
  \includegraphics[scale=0.18]{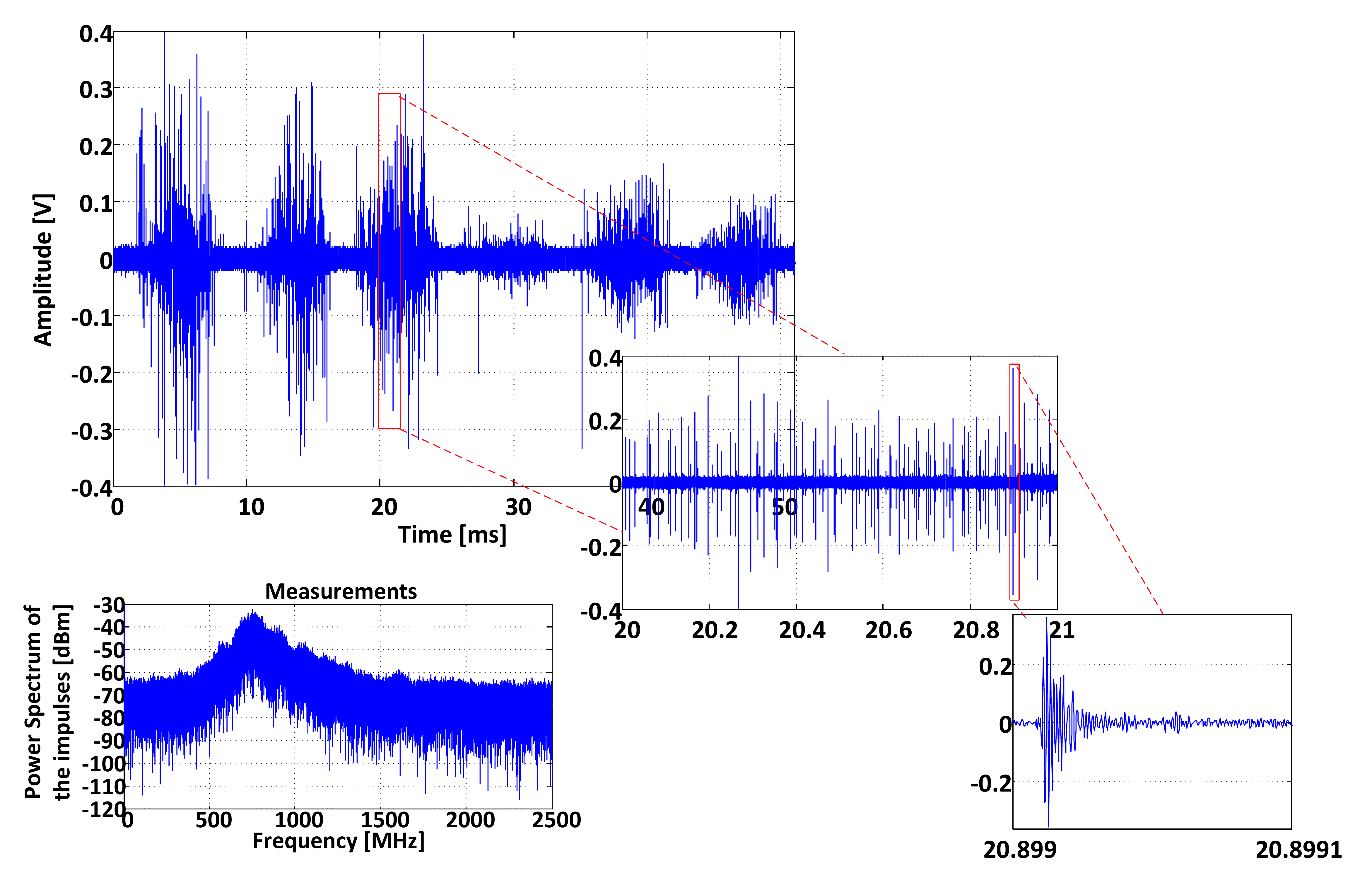}\\
  \caption{Illustrative impulsive noise measurement in a 735 kV substation, band: 800 MHz - 2.5 GHz; see~\cite{taiwan,cigre} for detailed experimental setup and considerations.}\label{735}
\end{figure}
Another particularity observed during the measurement campaign is that the impulses are often grouped every 8.3 ms (Figure~\ref{735}), which represents one half of a 60 Hz cycle. As we have explained previously, the partial discharges that produce impulsive noise, are created as soon as the electrical potential in the 60 Hz cycle reaches a critical value of the insulator; this critical value can be reached several times during a cycle and it is more likely that partial discharges occur around the maxima (positive and negative) of the 60 Hz cycle~\cite{bart}.\

The existing models are inappropriate to be used in wide band, because they cannot provide the appropriate correlation between the samples and a power spectrum of the impulses that is similar to the measurements performed in substations. The wide band model must have some memory to ensure a correlation between the samples; the Markov chain is an appropriate statistical tool that can implement the switching process and the varying shape of the impulses.
\section{Partitioned Markov chain model}

\subsection{Classic partitioned Markov chain}
\begin{figure}[h]
\begin{center}
  \includegraphics[scale=0.4]{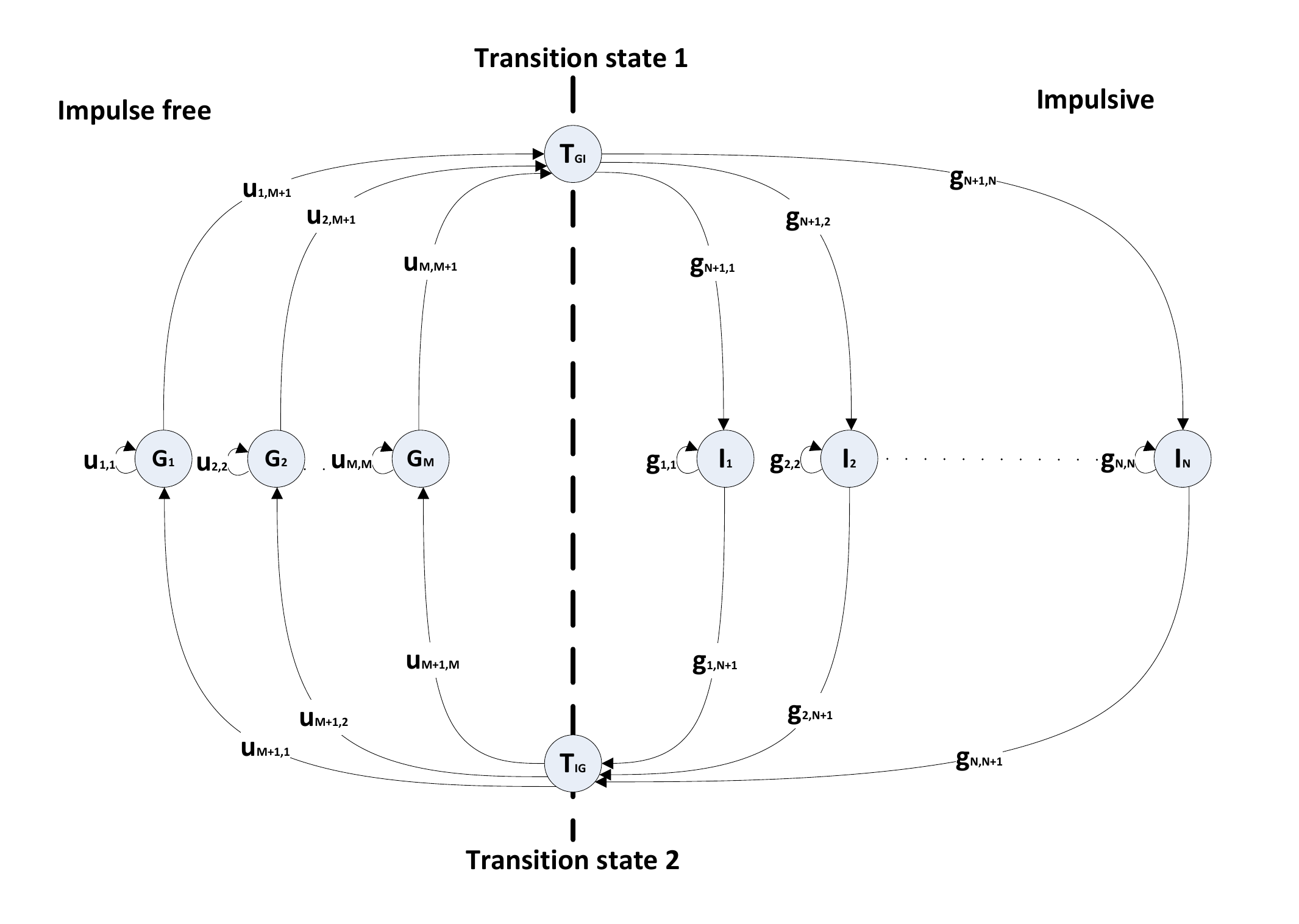}\\
\end{center}
  \caption{Zimmermann partitioned Markov chain model~\cite{zimmermann}.}\label{zimm}
\end{figure}

\begin{figure}[h]
\begin{center}
  \includegraphics[scale=0.32]{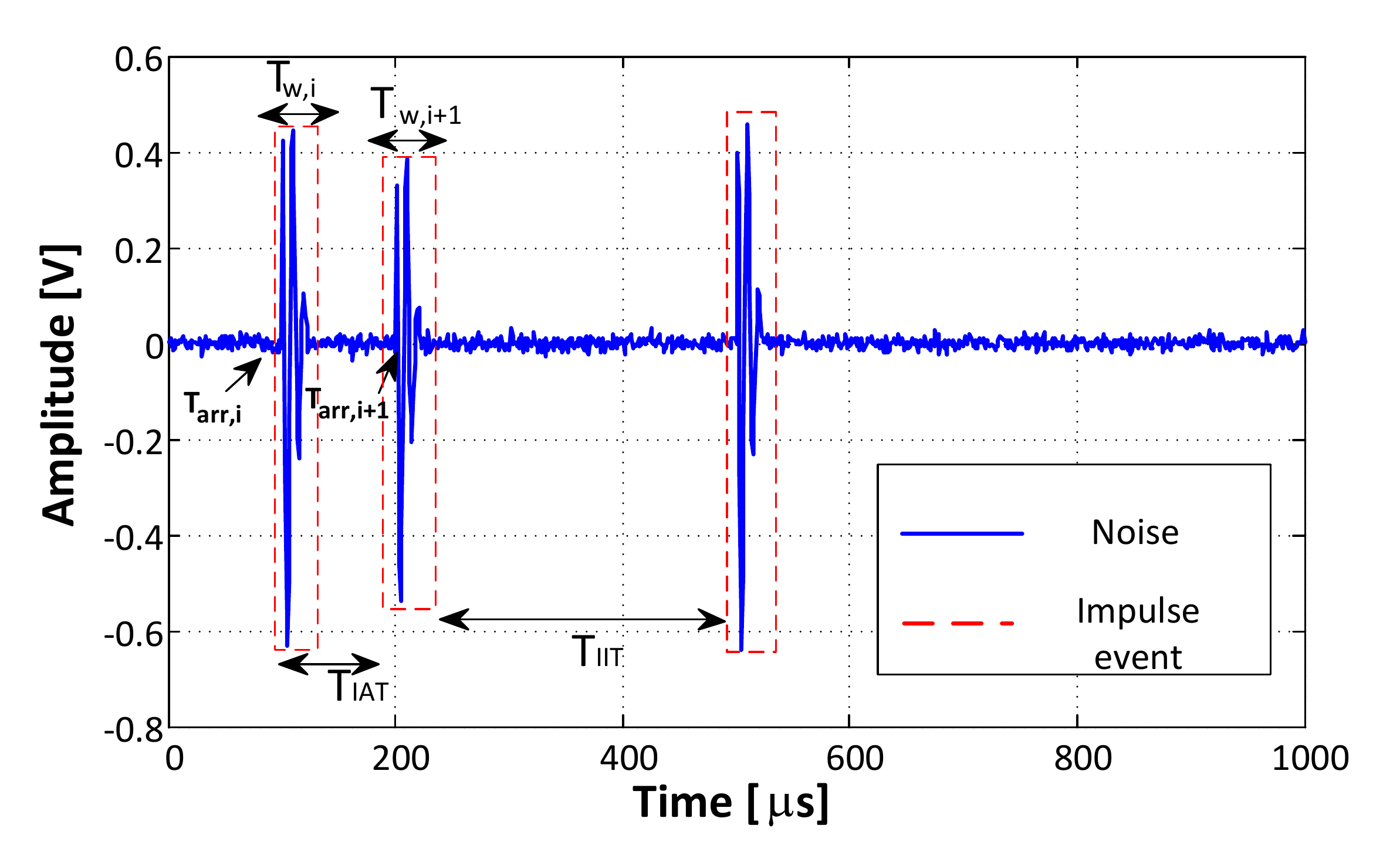}\\
\end{center}
  \caption{Detailed representation of impulse occurrence.}\label{TIAT}
\end{figure}

Existing partitioned Markov chains representing impulse events for Power Line Communications (PLC)~\cite{zimmermann} are composed of two groups of states, respectively the impulse-free states and the impulse states (Figure~\ref{zimm}); the impulse-free states represent the absence of impulses when only the background noise is observable and the impulse states represent the impulse events. Zimmermann~\cite{zimmermann} considers that the total noise is composed of $M$ background noise sources and $N$ impulse sources (Figure~\ref{zimm}) and that each state generates its own amplitude for a duration depending on the probability to remain in the state. Two transient states organize the transition from the impulse-free states to the impulse states and vice versa.\\
With the probability to remain in an impulsive state, Zimmermann attempts to control the impulse duration $T_w$ (Figure~\ref{TIAT}) and with the probability to remain in the different background noise states, he influences the impulses generation period $T_{IAT}$ (Figure~\ref{TIAT}). We call inter-arrival time (IAT) the duration between two impulse occurrences and inter-impulse time (IIT) the duration between two impulses (Figure~\ref{TIAT}).\

The impulsive noise observed by Zimmermann reveals that the inter-arrival times distribution seems to be a mixture of exponential distributions and for this reason, the model uses different states for the background noise.
The samples of the impulses are not generated from statistical distributions to emulate the shape of the observed impulses; only the impulse occurrence (Figure~\ref{TIAT}) is represented by samples equal to '1' in impulse states and the background noise is represented by samples equal to '0' in impulse-free states. Although Zimmermann observed that the amplitude of the impulses follows an exponential distribution~\cite{zimmermann}, no methods are proposed to replicate the sample distribution. This partitioned Markov chain provides satisfying results regarding the impulsive events, however the lack of samples generation is insufficient for a wide band representation of the communication channel. We propose then to modify the partitioned Markov chain by adding extra states, which can generate samples from Gaussian distributions.
\subsection{Proposed Model}
For a better understanding of our work, we explain first, based on substation measurements, how we interpret the events and the characteristics of the impulses; thereafter we present the configuration of our partitioned Markov chain.\

\begin{figure}[h]
\begin{center}
  \includegraphics[scale=0.15]{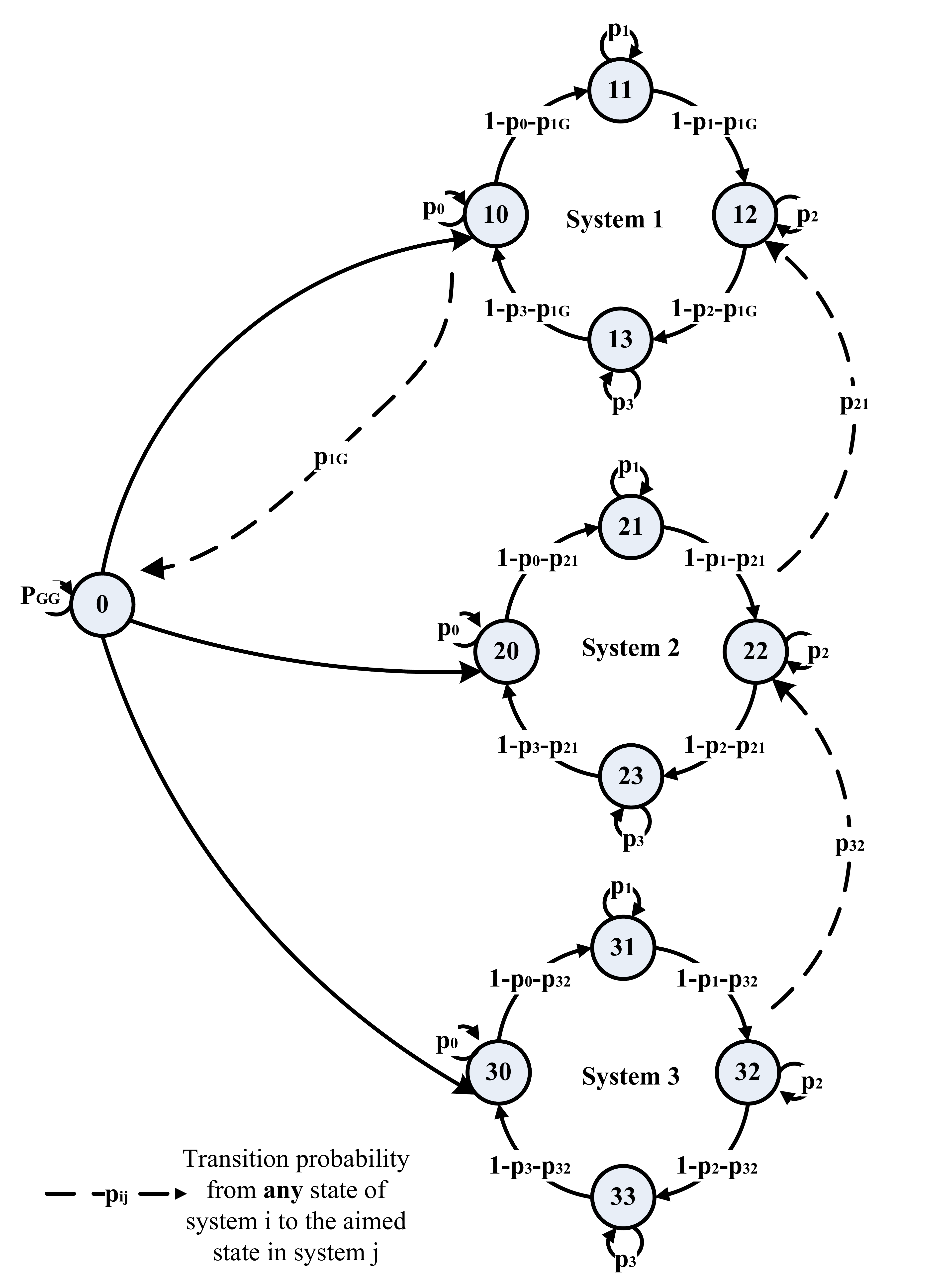}\\
\end{center}
  \caption{Proposed model: partitioned Markov chain with systems of 4 states configured to generate the impulses samples with an oscillating waveform.}\label{mc2}
\end{figure}

The background noise is constantly present in substations, however it is no longer observable when an impulse occurs. An impulse is observable as soon as its amplitude\footnote{The impulse amplitude is always positive and corresponds to the maximum of the absolute values of the samples in an impulse.} is superior to the background noise envelope, so we can interpret an impulse as a noise whose samples have a larger variance than the background noise. Our model is only based on observation, thus we consider that impulsive noise is a process switching from background noise to impulses and vice versa and we ensure that the variance of the samples corresponding to the impulses is superior to the background noise variance. Secondly, the impulses occur independently from each other, which means that the occurrence times are i.i.d. Moreover, as soon as an impulse occurs, the samples contained in the impulse are correlated to produce a damped oscillation waveform. Finally, we assume that there are three groups of impulses that gather the short, the medium and the long impulses, which are respectively the small, the medium and the large impulses in amplitude~\cite{cigre}. We introduce the ``impulsive state'' term as being a state of the partitioned Markov that does not represent the background noise (any state but state 0 in Figure~\ref{mc2}).
The samples of each group of impulses are implemented using a configuration of impulsive states, called ``impulsive system'' (Figures~\ref{mc2} and~\ref{OSC46}), where the mean and the variance of the samples are estimated from the impulses of the groups observed. For example, system $i$ aims to represent a group $i$ having an average amplitude $m_i$. The damped effect is ensured by the succession of systems with decreasing average amplitude $m_i$, $m_3>m_2>m_1$ (Figure~\ref{mc2}). For example, the impulses in group 3 (larger amplitude and longer duration), are generated by transition between states in system 3, then in system 2 and finally in system 1. With the sum of the time spent in each system being equal to the duration of the impulse in group $i$, we can ensure to provide the damping effect with the appropriate duration of the impulses.
\subsection{Implementation of damped oscillation}
\begin{figure}
  \includegraphics[scale=0.2]{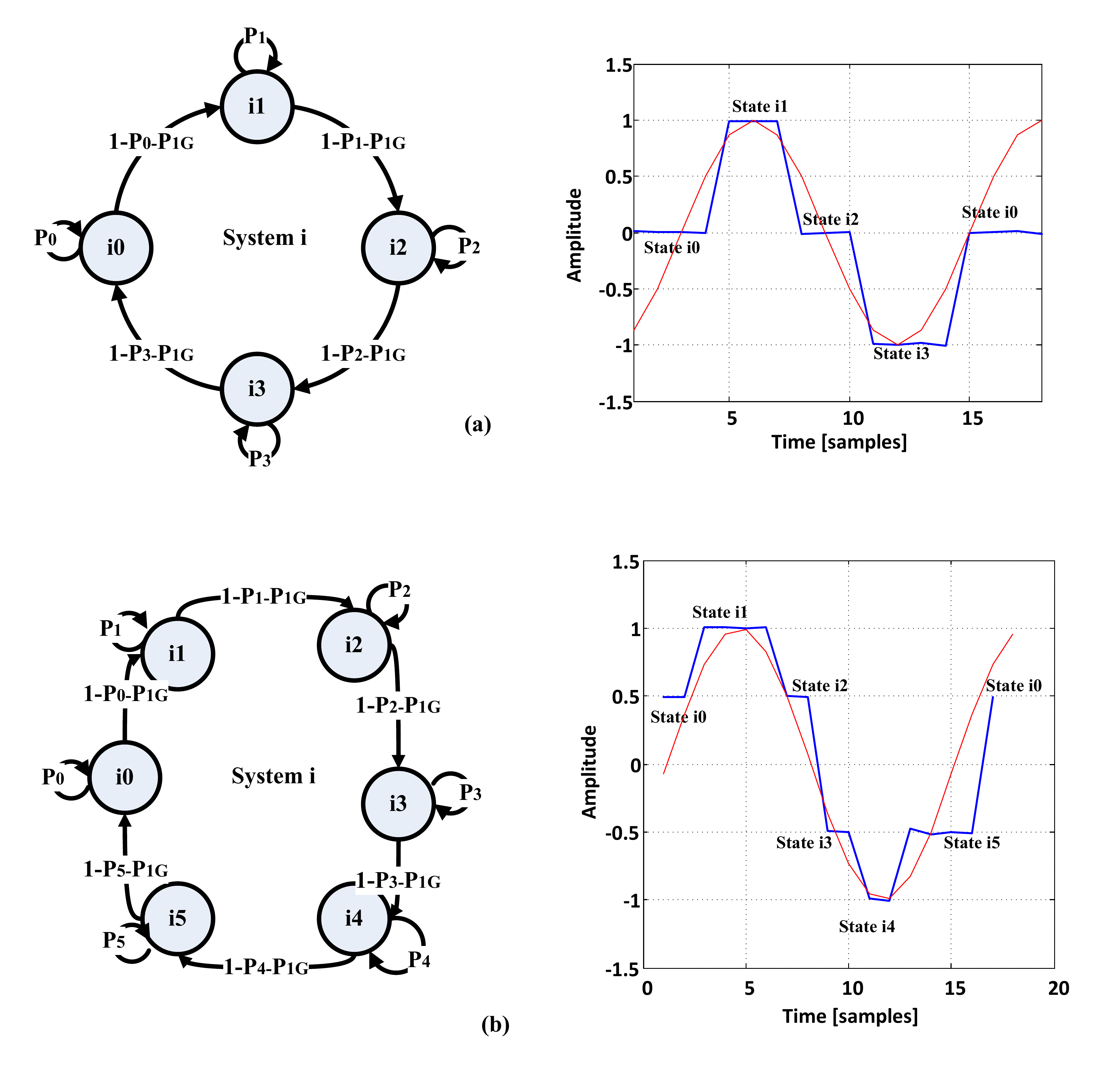}\\
  \caption{Implementation of an oscillation with 4 states (a) and 6 states (b) in an impulsive system $i$.}\label{OSC46}
\end{figure}

The impulsive systems that we have introduced in section III.B provide some oscillations within the impulse shape. The oscillations are implemented by each impulsive system, by using 4 or even 6 states (Figure.~\ref{OSC46}) in order to generate samples around values characterizing a sinusoidal signal. In an impulsive system, we call a ``loop'' the path of states when the process leaves an initial state, hits all other states in the system and returns to the initial state. For example, in figure~\ref{OSC46}, the paths of states $\{i0,i1,i2,i3,i0\}$ and $\{i0,i1,i2,i3,i4,i5,i0\}$ are ``loops''. The time spent in a loop corresponds to a sinusoidal period that we evaluate from the the oscillation frequency of the observed impulses. The number of states is inferior to the ratio of the sampling frequency to the desired frequency, because each state generates at least 1 sample; therefore the number of states is the minimum number of samples for one period of a sinusoid. For example, at a sampling frequency of 5 GS/s sinusoidal signals of frequency up to 1.25 GHz can be generated with 4 states. According to the spectrum of the measurements from Hydro-Quebec substations (Figure~\ref{735}), the highest frequency peak is centered at 800 MHz, for which 4 or 6 states are enough.

The generation of the samples value in impulsive systems is designed as follows:\\
For oscillations using 4 states :
\begin{itemize}
  \item state $i0$ :  $\mathcal{N}(0,\sigma^2_0)$
  \item state $i1$ :  $\mathcal{N}(m_i,\sigma^2_i)$
  \item state $i2$ :  $\mathcal{N}(0,\sigma^2_0)$
  \item state $i3$ :  $\mathcal{N}(-m_i,\sigma^2_i)$
\end{itemize}
For oscillations using 6 states :
\begin{itemize}
  \item state $i0$ :  $\mathcal{N}(m_i/2,\sigma^2_0)$
  \item state $i1$ :  $\mathcal{N}(m_i,\sigma^2_i)$
  \item state $i2$ :  $\mathcal{N}(m_i/2,\sigma^2_0)$
  \item state $i3$ :  $\mathcal{N}(-m_i/2,\sigma^2_0)$
  \item state $i4$ :  $\mathcal{N}(-m_i,\sigma^2_i)$
  \item state $i5$ :  $\mathcal{N}(-m_i/2,\sigma^2_0)$
\end{itemize}
where $\mathcal{N}(m,\sigma^2)$ represents the Gaussian distribution with mean $m$ and variance $\sigma^2$. $\sigma_0^2$ and $\sigma_i^2$ are respectively the variances of the background noise and the impulse amplitude of the group $i$, and $m_i$ is the mean of the impulse amplitude in group $i$.
\subsection{Sample values generation and impulse amplitude}
The samples values and the impulses amplitude are related because the amplitude of the impulse corresponds to the largest absolute value of all the samples generated by the impulsive state of a system. The impulse amplitude distribution is different depending on the application of the existing models because the samples are collected in different environments with different measurement setups. In the case of impulsive noise in PLC, Zimmermann~\cite{zimmermann} observed that the impulse amplitude seems to follow an exponential distribution, however he did not configure his partitioned Markov chain to generate the amplitude.\

Although Middleton models (Class A, B and C) do not replicate the time-correlated nature of impulsive noise, the probability density function (PDF) of the samples value is very often used in communication studies~\cite{midd1,midd2,midd3}. In Middleton models, the samples of the impulses is considered to be generated by an infinite number of Gaussian sources; therefore the Middleton class-A model can be interpreted as different states with different Gaussian distributions, thus we choose to represent the impulse samples with Gaussian distributions in our model. Another reason for using a Gaussian distribution for the samples generation is to simplify the parameter estimation. A Gaussian distribution requires only two parameters to generate samples, the mean and the variance, and for a maximum likelihood approach, we will only require the sample mean and the sample variance.
\subsection{Transition matrix}
The transition matrix parameterizes the events of impulsive noise such as impulse occurrences, impulse durations and oscillations. As we have explained, the oscillations are performed with the ``loop'' in impulsive systems and we set the frequency by selecting the appropriate probability to remain in each state. The diagonal coefficients of the matrix are all the same and they are estimated from the largest frequency peak of the impulse spectrum. Only the first diagonal coefficient is different from the others because it represents the probability to remain in the background noise and it is estimated from the number of impulses that are observed.\

This matrix is designed to create appropriate time correlations between the impulsive noise samples. Also it configures the occurrence times of the impulses with the probabilities to go from the state 0 to the states 10, 20 and 30 (Figure~\ref{mc2}). Basically, the transition matrix is set by using the information extracted from the measurements. As soon as we have detected the three groups of impulses, we analyze the mean duration of the impulses and the oscillation frequency. With the oscillation frequency we can set the probability to remain in each state of an impulsive system and to simplify the parameter estimation, we will take the same probability to remain in impulsive state for each system. Moreover, we configure the time spent in a system by setting the probability to go from the system $i$ to the system $i-1$ (Figure~\ref{mc2}). We will see in the parameter estimation section how to calculate the transition probabilities.
\section{Parameter estimation}
The goal of this section is to describe methods to determine the values of the Markov chain parameters with which we can generate a sequence of samples that is similar to the noise samples measured in substations. The goal is to achieve similarity between the sequences in terms of the distributions of the amplitude, the duration, the inter-arrival time and the power spectrum of the impulses.\

Our approach in estimating the impulsive noise is different from Middleton~\cite{midd1} because we do not consider the samples being i.i.d for the parameter estimation; in this respect, we first identify samples of the measured noise that correspond to impulses, in order to proceed to the parameter estimation thereafter. Once all the impulses are detected, we still have to decide on criteria to classify the impulses into the three groups that our model represents by impulsive systems. With the impulse samples classified in the three groups, it will be simpler to calculate the Markov chain transition probabilities and the parameters of the Gaussian distributions used to generate the samples in each state.
\subsection{Impulse detection from the measurements}
The aim of impulse detection is to classify all measured samples as either background noise or an impulse event. The background noise is often contained in an envelope with the value of the samples that does not exceed a threshold that we will call $th_a$.\

\begin{figure}
\begin{center}
  \includegraphics[scale=0.3]{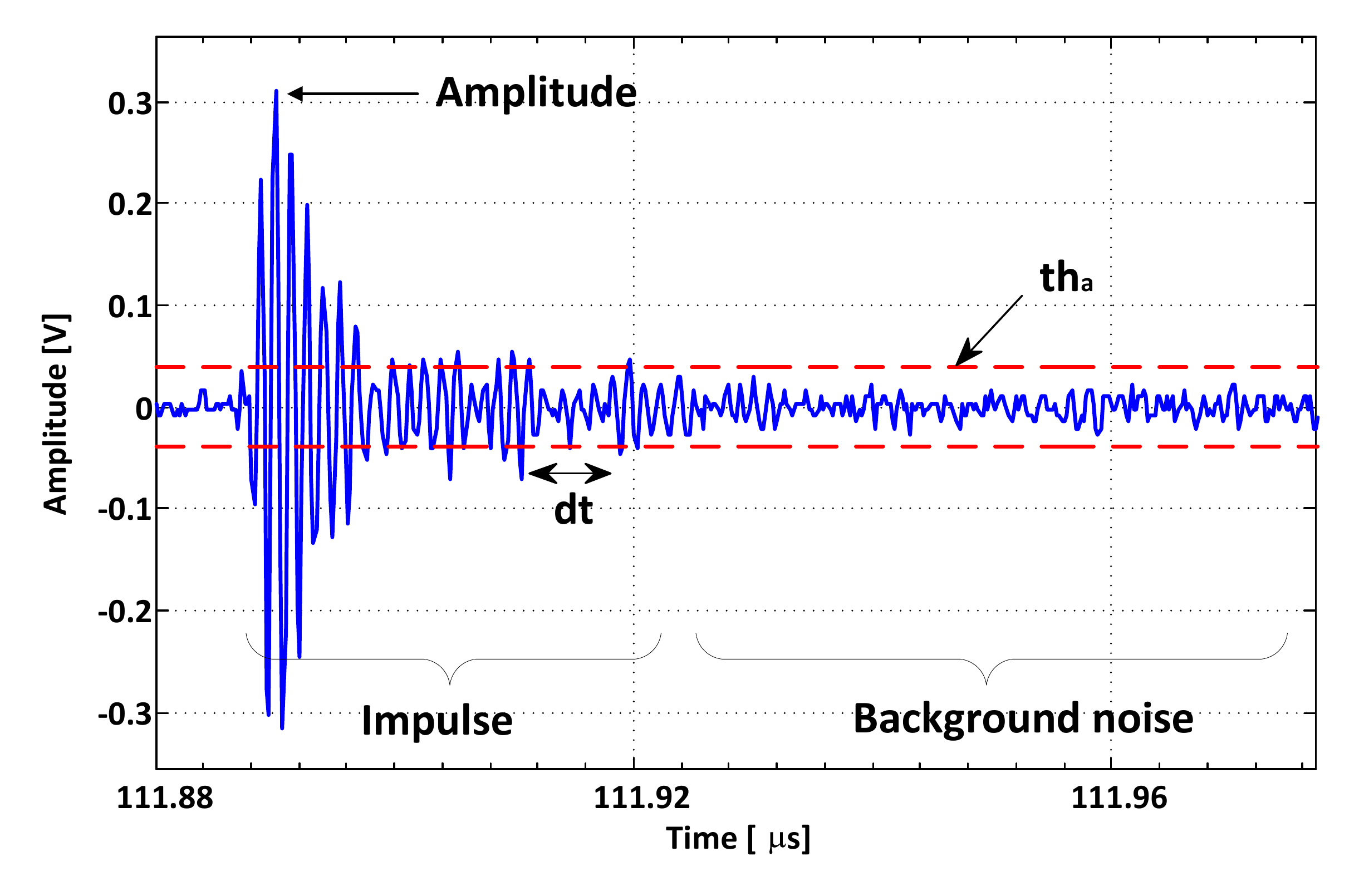}\\
\end{center}
  \caption{Analysis of an impulse based on the comparison of the samples with the background noise level.}\label{th}
\end{figure}

A first step is to find this threshold $th_a$ that separates the impulse samples from the background noise. All samples above this threshold, in absolute value, are deemed to belong to an impulse; however a part of the impulse is likely to have its samples value remaining for some period in the background noise envelope (Figure~\ref{th}); therefore the samples that belong to an impulse and that are also under the threshold $th_a$ must be considered by our method of impulse detection. The part of an impulse that remains under the threshold $th_a$ is composed of several consecutive samples, so as the background noise between two impulses, and the second step of the impulse detection will be to determine if those consecutive samples belong to an impulse or not; this decision will be based on the number of these consecutive samples compared to a duration threshold $th_d$, as described below.\

For the impulse detection, we will proceed as follows:
\begin{itemize}
  \item We determine a threshold $th_a$ that separates the background noise from the impulses (Figure~\ref{th}).
  \item We define the variable $dt$ as the duration between two consecutive samples above the threshold $th_a$.
  \item We collect the sequence $\{u_i\}$, which is the observation, in samples, of $dt$ from the measurements.
  \item We determine the threshold $th_d$ and for any element of $\{u_i\}$ above $th_d$, we consider that it corresponds to a duration between two different impulses.
  \item Once all the elements of $\{u_i\}$ are analyzed, we can separate the impulses from the background noise and we can extract the information about the amplitude, the duration and the inter-arrival time of the impulses.
\end{itemize}
We prefer evaluating the thresholds $th_a$ and $th_d$ by using as little prior information as possible because, when it will be embedded in communication devices, the system must be entirely autonomous faced with the environment.\

To find the threshold $th_a$ that separates the impulse samples from the background noise, we analyze the noise recording under an assumption of a simpler noise model. In order to simplify the calculation of $th_a$, we consider that the samples of the measured noise are distributed according to a Bernoulli-Gaussian process. If we would have considered all the states of our model, it would have been much more difficult to estimate the background noise variance; therefore we assume that the samples are distributed according to the pdf $f(x)$:
\begin{equation}\label{awgn}
    f(x)=\frac{\lambda}{\sqrt{2\pi\sigma_0^2}}\exp(-\frac{x^2}{2\sigma_0^2})+\frac{1-\lambda}{\sqrt{2\pi\sigma_1^2}}\exp(-\frac{x^2}{2\sigma_1^2}),
\end{equation}
where $\sigma^2_0$ is the background noise variance, $\sigma^2_1$ is the impulses samples variance and $\lambda$ is the probability to be in background noise.
By applying the method of moments~\cite{zabin1}, we calculate the three first even sample moments $e_2$, $e_4$ and $e_6$ and solve the system of equation where the sample moments are equal to the statistical moments to find the three parameters $\sigma_0^2$, $\sigma_1^2$ and $\lambda$. We define $\langle . \rangle $ as being the mean function and we calculate the three first even moments :
      $$
\left\{
    \begin{array}{lll}
        \langle x^2 \rangle = \lambda\sigma_0^2+(1-\lambda)\sigma_1^2  \\
        \langle x^4 \rangle = 3[\lambda{(\sigma_0^2)}^2+(1-\lambda){(\sigma_1^2)}^2] \\
        \langle x^6 \rangle = 15[\lambda {(\sigma_0^2)}^3+(1-\lambda){(\sigma_1^2)}^3]
    \end{array}
\right.
$$
The solutions of the system of equations $\langle x^i \rangle=e_i $, $i\in\{2,4,6\}$, give the following background noise variance $\sigma_0^2$ :

\begin{equation}\label{varg}
    \sigma_0^2=\frac{\alpha + \beta}{\gamma}
\end{equation}
where
      $$
\left\{
    \begin{array}{lll}
        \alpha=15e_2e_4-3e_6  \\
        \beta=\sqrt{75e_4^2(4e_4-9e_2^2)+270e_2(2e_2^2-e_4)e_6+9e_6^2} \\
        \gamma=90e_2^2-30e_4
    \end{array}
\right.
$$
With the variance estimated, we set the background noise level using the universal threshold~\cite{univ} $\sigma \sqrt{2\log n}$, with $\sigma$ being the standard deviation of a zero-mean Gaussian sequence of length $n$. Here, an appropriate sequence length $n$ would be an average inter-impulse time and according to several measurements, it would correspond to a value between $10^5$ and $10^6$ samples. Hence we use a background noise threshold $th_a=5\ \sigma_0$ and all samples above this threshold are considered to be part of an impulse.\

\begin{figure}[h]
\begin{center}
  \includegraphics[scale=0.2]{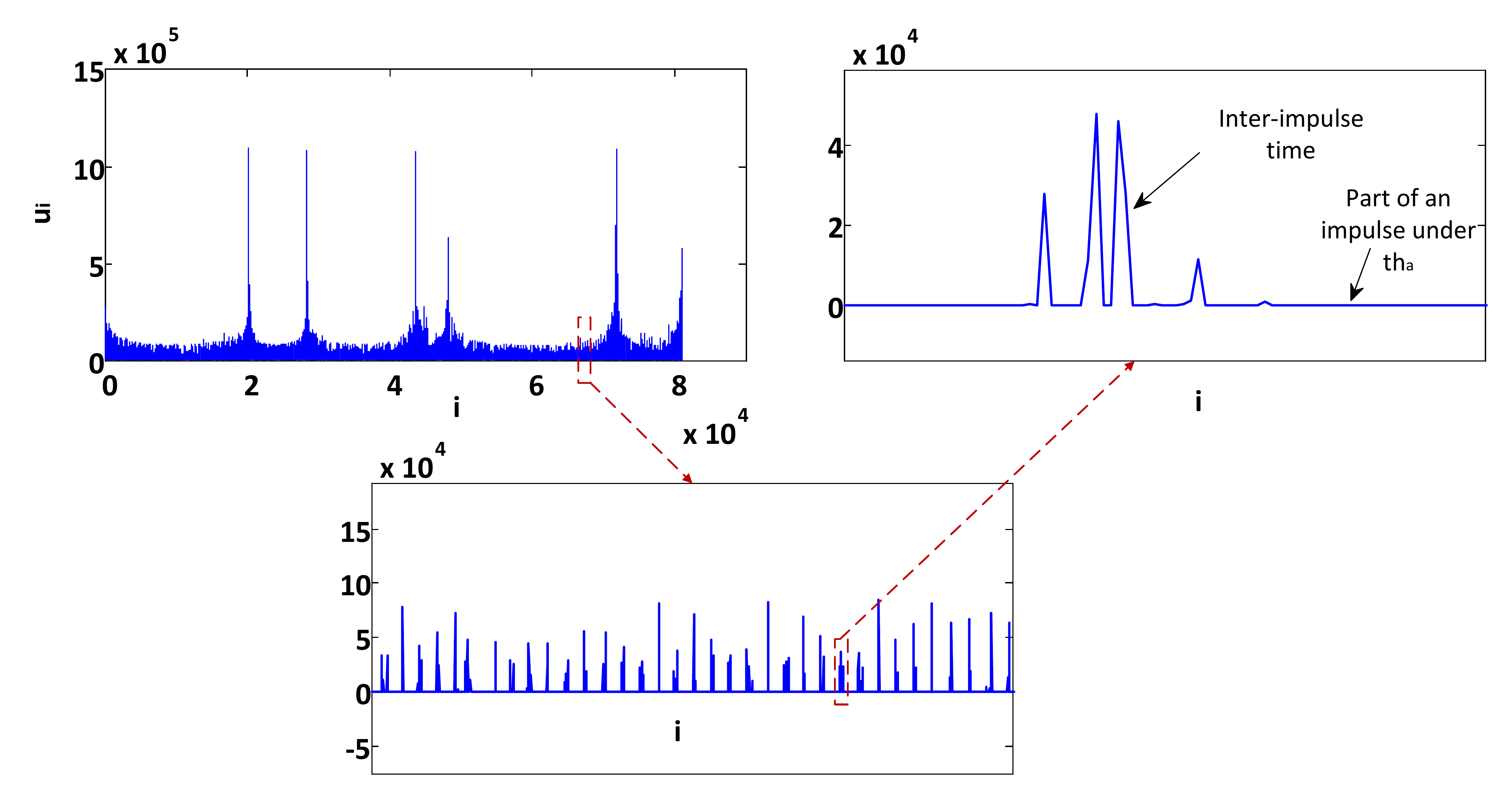}\\
\end{center}
  \caption{$\{u_i\}$ sequence calculated from measurements.}\label{DT}
\end{figure}

With the threshold $th_a$, we isolate the noise samples above the background noise level and we compose the sequence $\{u_i\}$ (Figure~\ref{DT}). Among the values calculated, some are a part of an impulse duration and the others correspond to inter-impulse times. Now the objective is to find the threshold $th_d$ that will separate the two kinds of duration.
After observation of $\{u_i\}$ sequences from different measurements, it seems that the elements of $\{u_i\}$ corresponding to a part of an impulse duration are an interval of integers, while the values of the inter-impulse times are all the rest with values estimated between 100 and 1 millions of samples~\cite{cigre}. We consider that the threshold $th_d$ is an integer value that breaks the continuity of the $\{u_i\}$ values, therefore we consider that the threshold $th_d$ is the smallest integer that does not belong to the $\{u_i\}$ sequence. The detection of the $\{u_i\}$ values that correspond to an inter-impulse time allows the separation between the background noise samples and the impulse samples.\

After extracting the impulses, we can find their amplitudes and their duration; moreover, we can create a signal composed of a succession of impulses and calculate its power spectrum.
\subsection{Sample value estimation}
The samples of each impulsive state are generated from Gaussian distribution with mean and variance estimated from the amplitude of the impulses. For the parameter estimation, we assume that the amplitude of the observed impulses are Gaussian, then using the maximum likelihood approach, for each impulse group $i$, we estimate the mean $m_i$ and the variance $\sigma^2_i$ by calculating respectively the sample mean and the sample variance of the sequence of amplitudes observed in each group.\

Each state provides samples using a Gaussian distribution with the appropriate mean and variance estimated from the impulse group and only the largest absolute value of the samples generated is considered as being the impulse amplitude for an impulse. In each impulsive system $i$, the two states using the means $m_i$ or $-m_i$ are more likely to provide the amplitude associated to a group of impulses.
\subsection{Groups of impulses}
\begin{figure}[h]
\begin{center}
    \includegraphics[scale=0.3]{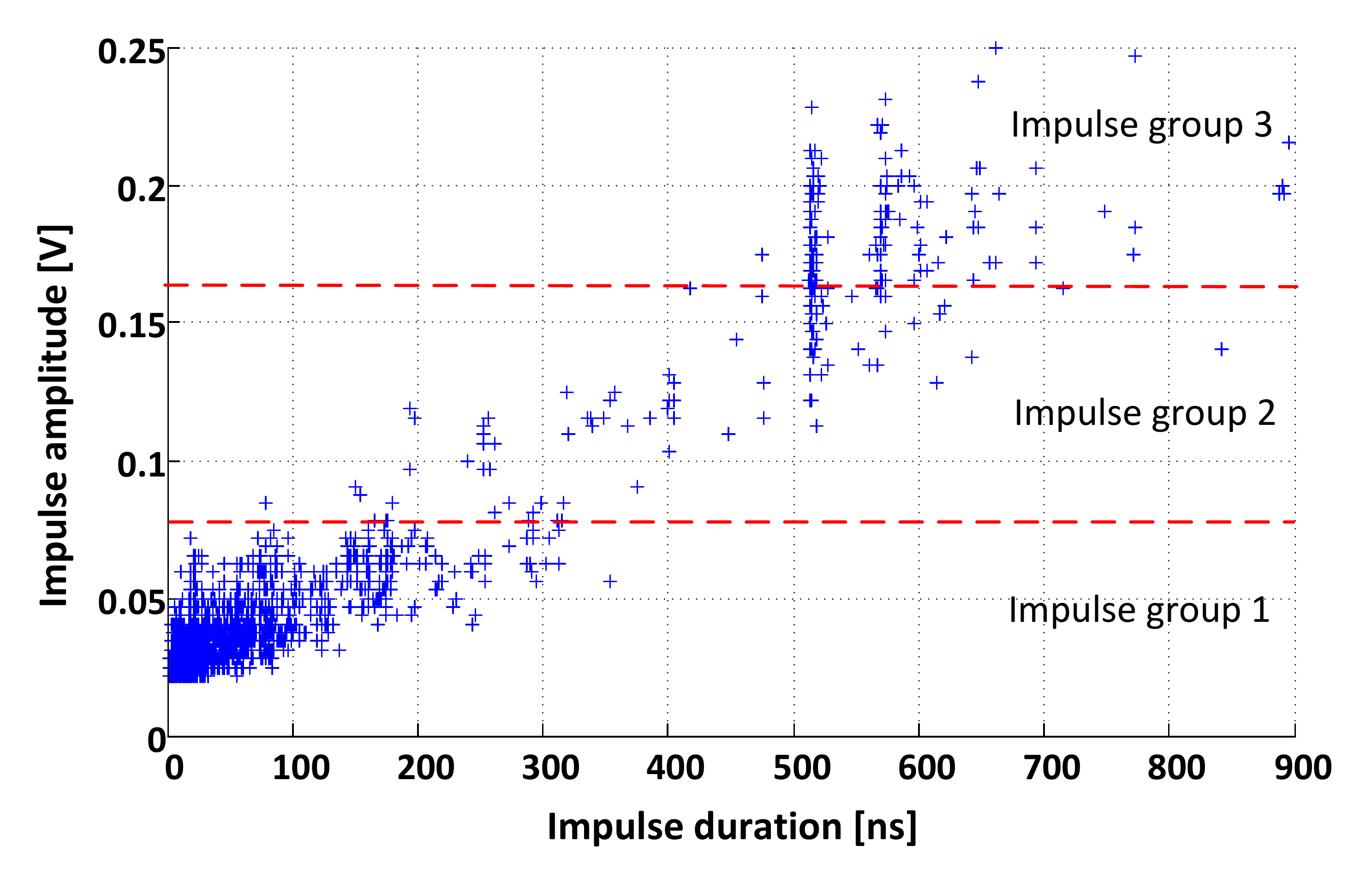}
\end{center}
  \caption{Two-dimensional plane of the amplitude vs the duration of the impulses.}\label{scatter}
\end{figure}

Once the impulses are separated from the background noise, it is now easier to analyze the groups of impulses in order to estimate the parameters required by the Markov chain. The impulse amplitude and the impulse duration are plotted on a two-dimensional plane to offer a better overview of the situation (Figure~\ref{scatter}). The belonging of an impulse to a group is determined by its amplitude, so to simplify the parameter estimation method, we split the amplitude sequence into three intervals with the same length in amplitude (Figure~\ref{scatter}).
Each group contains the information that will be used for the calculation of Markov chain transition probabilities and for the estimation of the parameters $m_i$ and $\sigma^2_i$ used in impulsive system $i$.
\subsection{Estimation of the transition matrix}
The occurrence of the impulses and their duration are implemented with the transition matrix of the Markov chain. The duration between the impulse occurrences (IAT) is configured by the probability to go from the background noise state to the impulsive system representing a group of impulses. With the impulse detection, we know how many impulses are distributed in each group, therefore we can find the probabilities required by the Markov chain. The maximum likelihood estimation of the transition probabilities is given by the following expression:

\begin{equation}\label{pij}
    p_{ij}=\frac{\mbox{number of transitions from $i$ to $j$}}{\mbox{number of samples in state $i$}}
\end{equation}
\

To configure the impulse duration, a first step is to find the average number of samples it takes for the Markov chain to leave an impulsive state. We just focus on state 0 for this example.\\
We calculate the probability to stay exactly $n$ instants in state 0 after starting in state 0 as:
\small
\begin{eqnarray}
Pr\{V_{0}=n\}&=& Pr\{X_{n}\neq0,X_{n-1}=0,\dots,X_1=0\ |\ X_0=0\}\nonumber\\
&=&(1-p_0)p_0^{n-1}. \label{eq1}
\end{eqnarray}
\normalsize
where $V_0$ is the time spent before leaving the state 0.\

Equation (\ref{eq1}) provides the distribution of the number of samples spent in state 0, and now, we are able to calculate the average time spent in state 0.\

We apply the following equation to calculate the mean time spent in state 0:
\begin{eqnarray}\label{EV00}
  E[V_{0}]&=&\sum\limits_{k=1}^{\infty}kPr\{V_{0}=k\}\nonumber \\
&=& (1-p_0)\sum\limits_{k=1}^{\infty}k p_0^{k-1}\nonumber \\
&=&\frac{1}{1-p_0}
\end{eqnarray}\

The duration spent in a state is configured with the probability to leave it, therefore the time spent in an impulsive system can be configured in the same way by using the probability to leave a system (i.e. $p_{32}$, $p_{21}$ and $p_{1G}$ in Figure~\ref{mc2}). If we consider an impulsive system as a single state, the probability to leave the system is the transition probability from any state of the system to another system or to the background noise state. To configure an average time T, in samples, spent in an impulsive system, the probability to leave each state of the impulsive system to another system, or to the background noise, is $p=1-\frac{1}{T}$. We estimate $T$ from the sample mean of the impulse duration of a group.
\subsection{Probability estimation for the oscillations}
To configure the oscillation frequency within the impulses, the model must have the appropriate probabilities to remain in the impulsive states in each system; with the probability to remain in an impulsive state, the model ensures an average sample duration for each state, which also contributes to ensure an average period of a sinusoid signal in an impulsive system (Figure~\ref{OSC46}). Intuitively, the average number of samples spent in a period ($T_{osc}$) of an impulsive system seems to be the sum of the average times spent in each state ($T_{osc}=\sum\limits_{i}\frac{1}{1-p_{i}}$). To verify this assumption, we calculate the average number of samples, $T_{osc}$ to return in a state after leaving it, in an impulsive system.\

\begin{figure}
\begin{center}
  \includegraphics[scale=0.3]{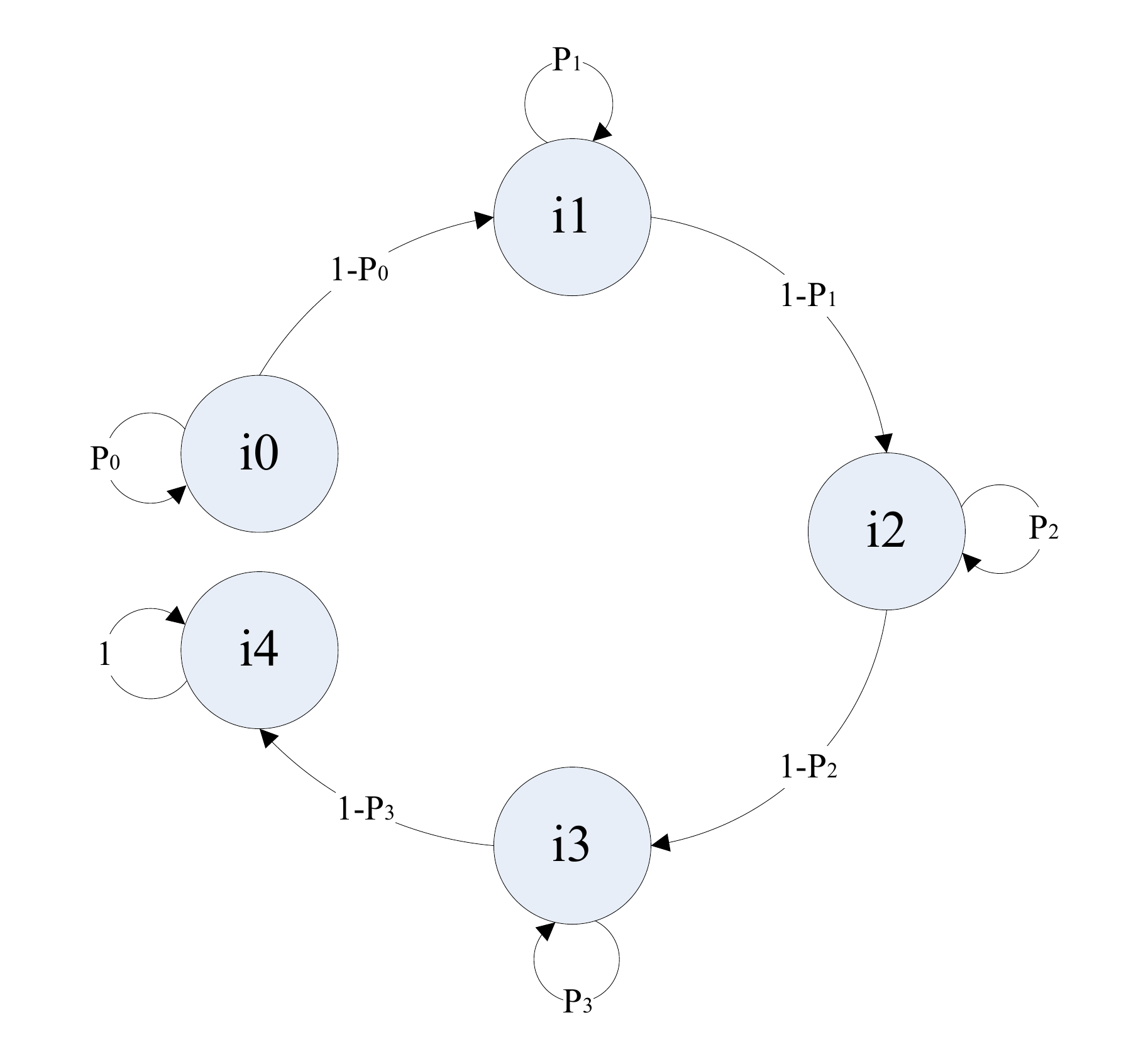}\\
  \caption{Modified impulsive system with an absorbing state and isolated from the partitioned Markov chain.}\label{MC3}
\end{center}
\end{figure}

We choose the example of a system of four states (Figure~\ref{OSC46}) to illustrate the concepts; the average time that we are interested in is the number of samples it takes for the process to return to the state $i0$ after leaving it. To study the average time of a ``loop'', we have to get rid of the other possible paths that might come after the transition from state $i3$ to state $i0$ and that do not characterize an oscillation period, therefore we modify the impulsive system in Figure~\ref{OSC46} by adding an absorbing state accessible after leaving the state $i3$ (Figure~\ref{MC3}). The absorbing state (state $i4$ in Figure~\ref{MC3}) represents any possible state that might come after returning to state $i0$. The average time of a loop, which is also the average period of the quasi-sinusoidal signal that the impulsive system attempts to replicate, is the average time before absorption when the initial state is $i0$. We calculate the average time to absorption~\cite{MC} by using the transition matrix of the Markov chain of an impulsive system.
For an absorbing Markov chain, the transition matrix has the following form:
\[
P=\left(
\begin{array}{c|c}
  \raisebox{-2pt}{{\Large\mbox{{$Q$}}}} \\[-3ex] & \vdots \\ \hline
   \dots&I\\
\end{array}
\right)
\]
\\
where $Q$ is the submatrix representing the process as long as it remains in transient states and $I$ is the identity matrix representing the absorbing states. Here, we have the transition matrix :
\begin{center}
$P=\left(
  \begin{array}{cccc|c}
    p_0 & 1-p_0 & 0 & 0 & 0 \\
    0 & p_1 & 1-p_1 & 0 & 0 \\
    0 & 0 & p_2 & 1-p_2 & 0 \\
    0 & 0 & 0 & p_3 & 1-p_3 \\ \hline
    0 & 0 & 0 & 0 & 1 \\
  \end{array}
\right)$
\end{center}\
\\
\begin{center}
$Q=\left(
  \begin{array}{cccc}
    p_0 & 1-p_0 & 0 & 0 \\
    0 & p_1 & 1-p_1 & 0 \\
    0 & 0 & p_2 & 1-p_2 \\
    0 & 0 & 0 & p_3 \\
  \end{array}
\right)$
\end{center}\
\\
We introduce the fundamental matrix: $N=(I-Q)^{-1}$~\cite{MC}
\\
\begin{center}
$N=\left(
  \begin{array}{cccc}
    \frac{1}{1-p_0} & \frac{1}{1-p_1} & \frac{1}{1-p_2} & \frac{1}{1-p_3} \\
    0 & \frac{1}{1-p_1} & \frac{1}{1-p_2} & \frac{1}{1-p_3} \\
    0 & 0 & \frac{1}{1-p_2} & \frac{1}{1-p_3} \\
    0 & 0 & 0 & \frac{1}{1-p_3} \\
  \end{array}
\right)$
\end{center}
 According to~\cite{MC}, the average time before absorption from any of the non-absorbing states can be found using the following vector $t$, which is the vector of the average times before absorption after leaving each state.\\
 \begin{eqnarray}
 t &=& \left(
  \begin{array}{cccc}
    \frac{1}{1-p_0} & \frac{1}{1-p_1} & \frac{1}{1-p_2} & \frac{1}{1-p_3} \\
    0 & \frac{1}{1-p_1} & \frac{1}{1-p_2} & \frac{1}{1-p_3} \\
    0 & 0 & \frac{1}{1-p_2} & \frac{1}{1-p_3} \\
    0 & 0 & 0 & \frac{1}{1-p_3} \\
  \end{array}
  \right)
    \times
 \left(
     \begin{array}{c}
       1 \\
       1 \\
       1 \\
       1 \\
     \end{array}
 \right)\nonumber\\
 \nonumber\\
  \nonumber\\
 &=& \left(
     \begin{array}{c}
       \frac{1}{1-p_0}+\frac{1}{1-p_1}+\frac{1}{1-p_2}+\frac{1}{1-p_3} \\
       \frac{1}{1-p_1}+\frac{1}{1-p_2}+\frac{1}{1-p_3} \\
       \frac{1}{1-p_2}+\frac{1}{1-p_3} \\
       \frac{1}{1-p_3} \\
     \end{array}
 \right)\nonumber\\
\end{eqnarray}
We are interested in the time spent in the Markov chain before absorption when the initial state is $0$, which corresponds to the first coefficient of vector $t$. The average time to absorption is $\frac{1}{1-p_0}+\frac{1}{1-p_1}+\frac{1}{1-p_2}+\frac{1}{1-p_3}$, which is the sum of the mean times spent in each non-absorbing state.\

To lower the complexity of the model, we choose the same transition probability for each pair of states in the impulsive systems. For a system using 6 states, the average time spent in a loop is $\frac{6}{1-p}$. For a given impulse frequency $f$ in $samples^{-1}$, each state in an impulsive system must have a probability $p_{i}=1-6f$. For the simulations, we will use impulsive systems with 6 states because this configuration offers a better implementation of the oscillations.
\section{Results and discussion}
With our partitioned Markov chain configured with 6 states per system, we generate the same number of samples as the sequence of measured noise (256 millions of samples), and by using our impulse detection method, we calculate the empirical distributions of the impulse duration, the impulse amplitude, the inter-arrival time and the whole noise samples (impulses and background noise). In order to evaluate the performances of our model, we study the divergence between the distribution of the measurements and our model for the impulsive noise characteristics. We propose to calculate the Kullback-Leibler divergence (KL-divergence) coefficient, which is a non-symmetric measure to estimate how close a distribution is to another~\cite{KL}; usually it measures the distance between an observation generated by a model that tries to be similar to another model. The KL-divergence coefficient over a set $X$ is given by : $coeff_{_{KL}} = \sum\limits_{x\in X}log(\frac{p(x)}{q(x)})p(x)$, where $p$ and $q$ are respectively the distribution of the impulsive noise characteristic coming from the measurements and the model. We also propose to calculate the Mean Square Error (MSE) of the cumulative density function (CDF) between the distributions coming from our model with the distributions coming from the measurements. For the same impulsive noise characteristics, we compare the divergence coefficients of the measurements from our model but also from the Bernoulli-Gaussian with memory~\cite{PLC} and the Middleton class-A~\cite{midd1} models.
\subsection{Measurement campaign in Hydro-Quebec substations}
The measurement setup is the same as in~\cite{taiwan,cigre}, i.e we use a wide-band antenna and a digital oscilloscope set with a 52.1 ms time window, 5 Giga-Samples per second as sampling frequency, which provides 256 millions noise samples. The measurements are performed to cover the 800 MHz - 2.5 GHz band, which contains the carrier frequencies of some of the classic wireless communications standards. The measurements take place within a 230 kV substation and the antenna is approximatively located 20 meters from any power equipment.
\subsection{Bernoulli Gaussian with memory and Middleton class-A models}
The Bernoulli-Gaussian model with memory~\cite{PLC} is a Markov chain with two states; it is a version of our model where all the impulsive states are replaced by one single state that generates samples with a zero-mean Gaussian distribution. The transition matrix is estimated using the same method as for our model. The probability to go from the background noise to the impulsive state is calculated from equation~(\ref{pij}) by using the number of impulses in the time window. The probability to remain in the impulsive state is calculated using equation~(\ref{EV00}) and the sampled mean of the impulses duration. The variances of the Gaussian distributions associated with the states are the sampled variance calculated from the samples belonging to the background noise and the impulses.\

The Middleton Class-A model has the following PDF expression:\

\begin{equation}\label{Mid_pdf}
        f(x)=\frac{e^{-A}}{\sqrt{2\pi}}\sum\limits^{\infty}_{m=0}\frac{A^m}{m!\sigma_m}\exp(-\frac{x^2}{2\sigma^2_m})
\end{equation}
with $\sigma_m^2=\sigma^2 \frac{m/A+\Gamma}{1+\Gamma}$ and $\sigma^2$ is the variance of the impulsive noise.\

The parameters $\Gamma$ and $A$ characterize respectively the impulse magnitude and their occurrence rate and they are estimated using the method of moments~\cite{midd3}.
\subsection{Model performance}
For the three models, we generate 256 millions samples and we analyze the noise sequences to extract the distributions of the impulsive noise characteristics. The Middleton PDF is used to generate i.i.d samples, then it only helps to compare the distribution of the noise samples values with the other models. According to Tables~\ref{KLD} and~\ref{MSE}, we observe that our model performs better than the Bernoulli-Gaussian with memory for the distributions of the impulses characteristics and quite similarly to Middleton for the distribution of the samples value.\

Both tables indicate that our model generates impulses with a duration distributed more similarly to the measurements than the Bernoulli-Gaussian model. Our model considers three groups of impulses while the Bernoulli-Gaussian model considers just one; hence we gain more precision and we manage to provide a more accurate distribution for the impulse duration.\

The inter-arrival time seems to be distributed equivalently for our model and the Bernoulli-Gaussian model with memory; the divergence coefficients are not so different between the models, which makes sense because the impulse occurrence is basically implemented in the same way for both models. However, our model offers better results because the process that generates the impulse occurrence is dependent of the process that generates the impulse duration. The partitioned Markov chain can only generate the first sample of an impulse when the current state is the background noise, then a more accurate implementation of the impulse duration will also provide a more accurate implementation of the inter-arrival time.\

While observing the total noise samples for the three models and for the measurements, we remark that the impulse samples are too few compared to the number of background noise samples (1 sample over $10^4$ belongs to an impulse according to the measurements), thus they are hardly observable on the empirical distributions. The Middleton Class-A gives a slightly better result for the KL-divergence, but the MSE of the CDFs for the three models is similar, therefore we can affirm that the samples value of our model is satisfyingly generated. Moreover, the assumption made for the estimation of the threshold $th_a$ is coherent because the divergence coefficients calculated between the measurements and our model are equal to the divergence coefficients calculated between the measurements and the Bernoulli-Gaussian model.\

\begin{figure}
  \includegraphics[scale=0.22]{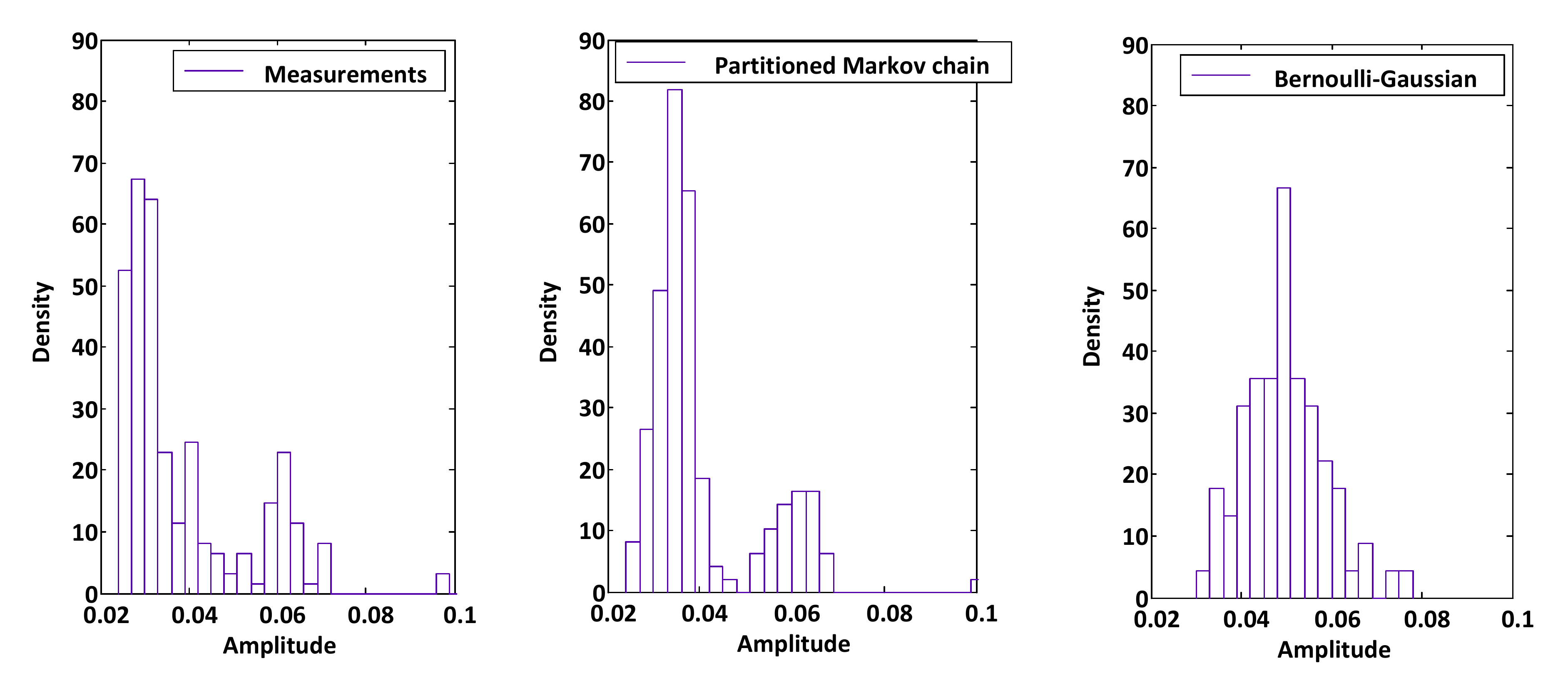}\\
  \caption{Distributions of the impulse amplitude for the measurements, the partitioned Markov chain and the Bernoulli-Gaussian model with memory.}\label{AMPCOMP}
\end{figure}

The three groups of impulses also give good results for the distribution of the impulse amplitude for our model. For the KL divergence and the MSE of the CDFs we have a more significant difference between our model and the Bernoulli-Gaussian. This difference can be explained by the implementation of the impulsive systems that uses Gaussian distributions with a non-zero mean and a variance, while the Bernoulli-Gaussian model only uses a zero-mean Gaussian distribution with a larger variance. This difference can be observed on the histograms of the impulse amplitude for both models and the measurements (Figure~\ref{AMPCOMP}). We distinguish the distribution of the three groups of amplitude observed from the measurements and from our model, however the Bernoulli-Gaussian model cannot provide such a precision.\

\begin{figure}
\includegraphics[scale=0.24]{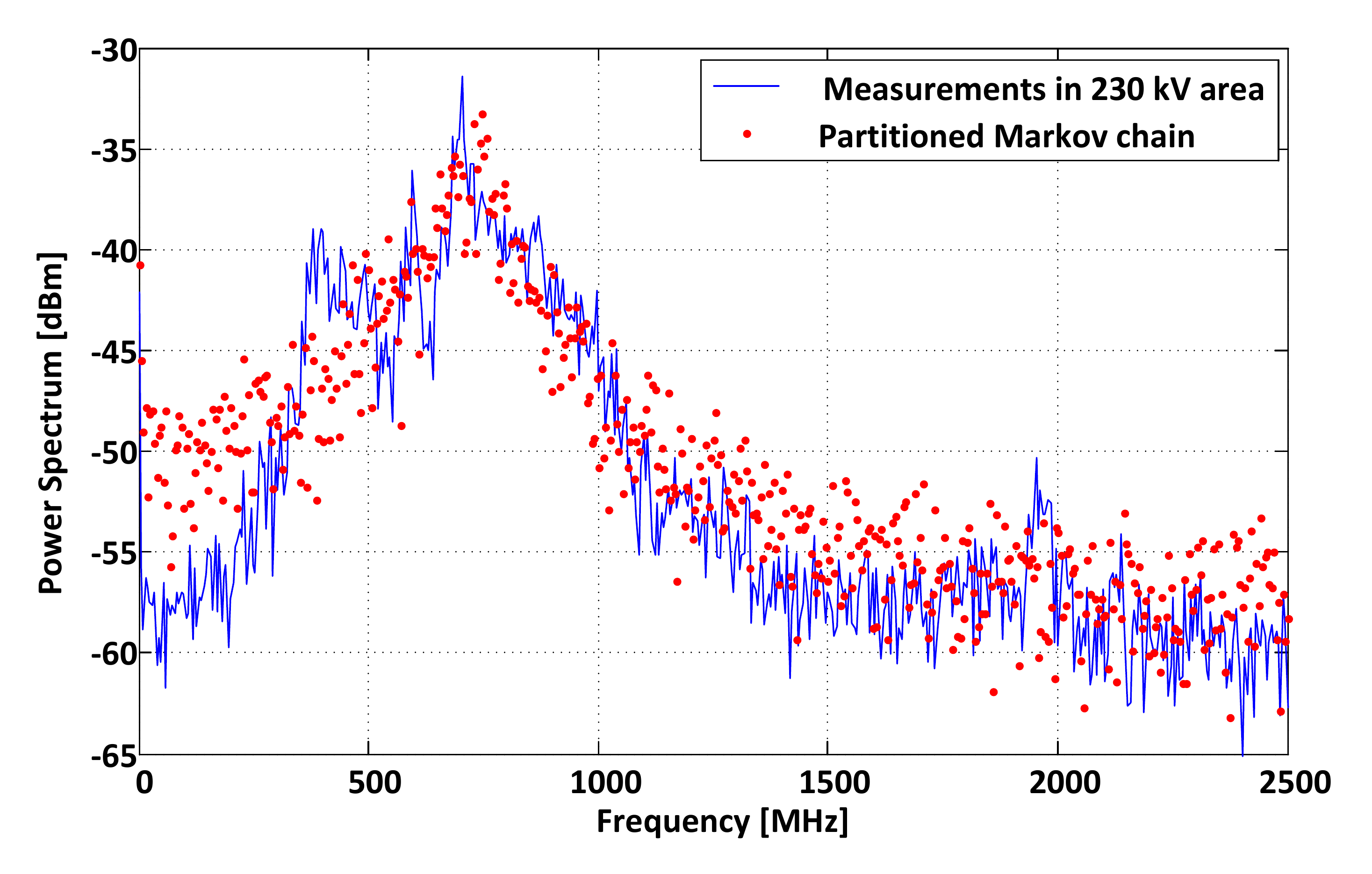}
  \caption{Power spectrum of noise generated by proposed system as compared to measurements.}\label{spcomp}
\end{figure}

\begin{center}
\begin{figure}
\includegraphics[scale=0.3]{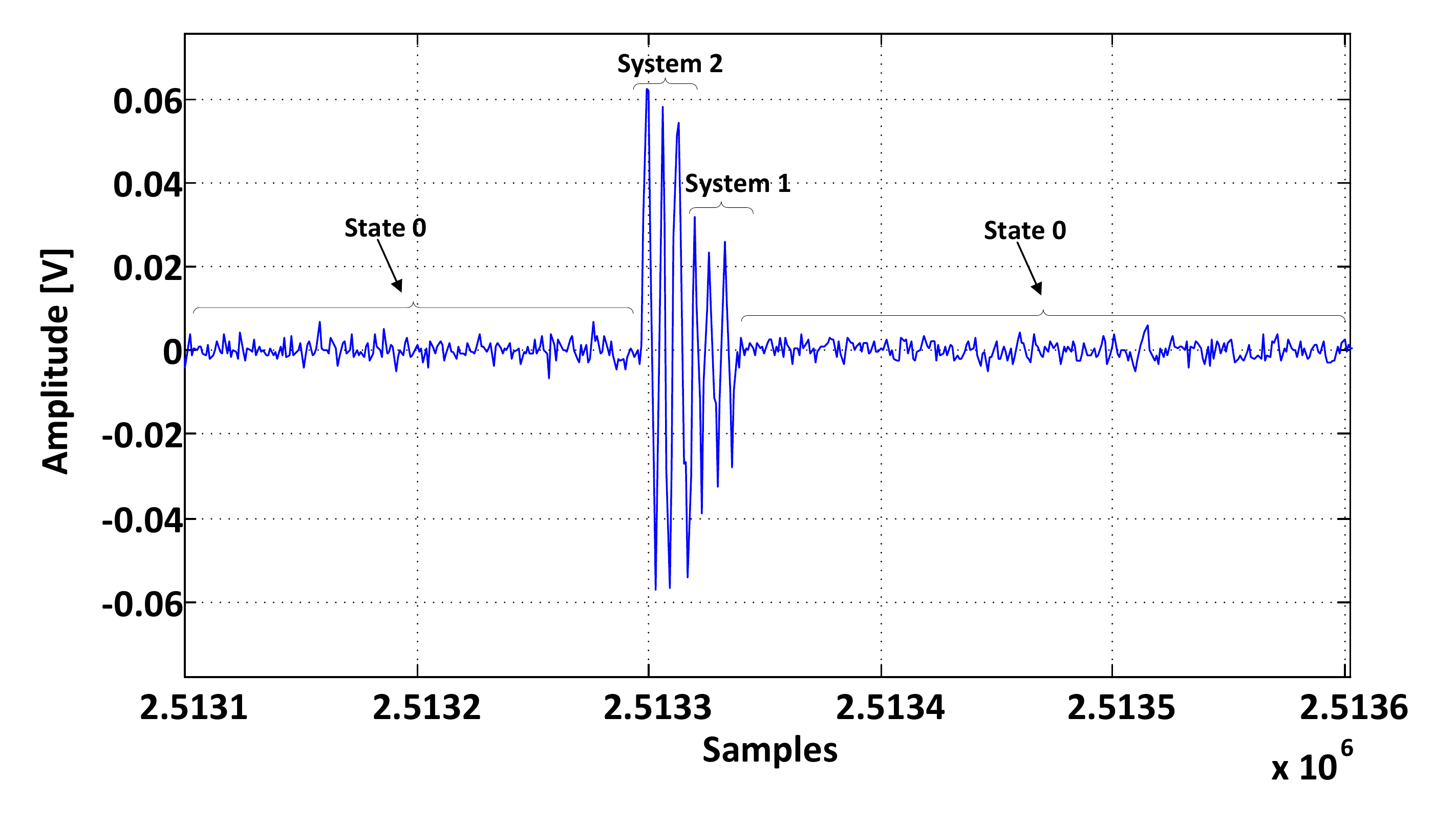}
  \caption{Example of an impulse generated by the Markov chain using 6 states per system.}\label{impulse}
\end{figure}
\end{center}

Another characteristic that is important for a wide-band representation of the noise is the power spectrum. We have gathered all the impulses from the measurements to calculate the average power spectrum and we did the same for the noise generated by our model (Figure~\ref{spcomp}). When we focus around the classic wireless frequencies, respectively 900 MHz, 1.8 GHz, and 2.4 GHz, our model is, in average, 3 dB from the measurements. The other models cannot provide such a power spectrum, because they do not provide the appropriate correlation between the samples within the impulses.
Finally, while observing an impulse generated by our model, we remark that the damped effect is well performed (Figure.~\ref{impulse}). We can also distinguish the parts of the impulse that have been generated by the impulsive systems of our model.

\begin{table}
  \centering
  \begin{tabular}{|c|c|c|c|c|}
   \hline
    \centering{$-$} & Class-A & BG with memory & MC Model \\\hline
    \centering{Samples value} & \textbf{0.0679} & 0.0808 & 0.0804 \\\hline
    \centering{Impulse duration} & \centering{$-$}  & 0.6149 & \textbf{0.4325} \\\hline
    \centering{IAT} & \centering{$-$}  & 0.7582 & \textbf{0.6901} \\\hline
    \centering{Impulse amplitude} & \centering{$-$}  & 1.2102 & \textbf{0.7579} \\
    \hline
  \end{tabular}
  \caption{KL-Divergence of the impulsive noise models from the measurements for the distribution of the impulsive noise characteristics}\label{KLD}
\end{table}

\begin{table}
  \centering
  \begin{tabular}{|c|c|c|c|c|}
   \hline
    \centering{$-$} & Class-A & BG with memory & MC Model \\\hline
    \centering{Samples value} & \textbf{0.0125} & \textbf{0.0125} & \textbf{0.0125} \\\hline
    \centering{Impulse duration} & \centering{$-$}  & 2.1856 & \textbf{0.8563} \\\hline
    \centering{IAT} & \centering{$-$}  & 0.8722 & \textbf{0.6458} \\\hline
    \centering{Impulse amplitude} & \centering{$-$}  & 4.7873 & \textbf{0.7281} \\
    \hline
  \end{tabular}
  \caption{MSE of CDF for the impulsive noise characteristics between the measurements and the impulsive noise models}\label{MSE}
\end{table}

\subsection{Limits of the model}
Although the model usually generates noise samples similar to measurements, it might happen that some measurements are more difficult to represent. The difficulty generally concerns the replication of the spectrum. Some points need still to be improved, such as:\
\begin{itemize}
  \item The configuration of the Markov chain to generate the damping effect. The succession of impulsive systems with decreasing amplitude might provide the right amplitude and impulse duration on average, but it also modifies the waveform of the impulse. For example, when the model attempts to produce a medium impulse (group 2) using system 2 and system 1, while configuring the parameters, we have to adapt the time spent in system 2 considering the average time to be spent in system 1 and the mean duration of an impulse in group 2. Hence, the samples of system 2 would be generated for a duration that is not similar to the impulse observed in group 2. If the duration in system 2 is too long, it can increase the power spectrum of the impulses.
  \item The method used to detect the impulse groups should consider both the impulse duration and the impulse amplitude. The measurements in substation do not always provide a two-dimensional plane of the amplitude versus the impulse duration in which the points are linearly dispatched (Figure~\ref{scatter}). A clustering algorithm would give more representative groups of impulses, which could improve the distributions of the amplitude and the duration of the impulses that are generated by our model.
\end{itemize}
\section{Conclusion}
Although the Markov chain we propose is very complex, the results provided by the simulations are satisfying for a wide band representation. In time domain, the amplitude of the simulated impulses are distributed similarly to the impulses amplitude collected from measurements, as the distributions of the impulse duration and the impulse occurrence. The samples shaping the impulses ensure the damping and the oscillating effects, which provides the spectrum required by the wide band representation. Although this implementation requires 19 states, this partitioned Markov chain can be used by any wireless receiver with a carrier frequency between 800 MHz and 2.5 GHz.\

The partitioned Markov chain model requires many states to ensure the accuracy of the time-correlation of the noise. We prefer to use 6 states in the impulsive systems because it provides more accuracy when implementing the oscillations, which helps improving the similarity between the power spectrum of the impulses that are measured and the ones that are simulated.\

As future work we recommend to reconfigure the Markov chain in order to provide a more realistic pattern for the impulses. Ideally, every group of impulses should be represented by three impulsive systems in order to implement the damped oscillation effect with an appropriate rise and fall time. Moreover, in order to improve the model performances, the selection of the impulse groups must be improved by a clustering algorithm that automatically detects the centers and the radius of the clusters in the two-dimensional plane and gives their coordinates for configuring the amplitude and the duration of the impulses in the Markov chain. Finally, our model of Markov chain can be used to implement an optimum receiver for impulsive noise in substation. The receiver could detect the presence of impulses and based on the impulse samples information, it could correct and help for reinterpreting the interfered signal. A narrow-band version of the proposed model can also be mapped to study the communication under impulsive noise for different wireless systems.
%
%
%
\section*{Acknowledgment}
This work was supported by Hydro-Qu\'{e}bec, the Natural Sciences and
Engineering Research Council of Canada and McGill University in the
framework of the NSERC/Hydro-Qu\'{e}bec/McGill Industrial Research Chair
in Interactive Information Infrastructure for the Power Grid.

\ifCLASSOPTIONcaptionsoff
  \newpage
\fi



\bibliographystyle{IEEEtran}


\balance
\end{document}